\documentclass[12pt]{article}
\usepackage{graphicx}
\usepackage{amsmath,amsthm,amssymb,bm,subfigure,color,url,lineno,cite,setspace}
\usepackage[colorlinks=true,citecolor=amazonite,linkcolor=amazonite,urlcolor=amazonite]{hyperref}

\definecolor{amazonite}{RGB}{0,115,150}
\definecolor{myred}{RGB}{255,56,0}
\definecolor{mygreen}{RGB}{30,150,30}
\definecolor{mybrown}{RGB}{150,30,30}
\definecolor{darkblue}{RGB}{30,50,200}
\definecolor{darkred}{RGB}{80,30,30}
\definecolor{gray}{RGB}{180,180,180}

\graphicspath{{fig/}}

\begin{document}

\baselineskip 0.7cm

\renewcommand{\thefootnote}{\fnsymbol{footnote}}
\begin{flushright}
\end{flushright}

\vskip 1.35cm
\begin{center}
{\large \bf
Statistical Evolutionary Laws in Music Styles
}
\vskip 1.2cm

Eita Nakamura$^1$\footnote[1]{Electronic address: \tt{eita.nakamura@i.kyoto-u.ac.jp}}
and Kunihiko Kaneko$^2$\footnote[2]{Electronic address: \tt{kaneko@complex.c.u-tokyo.ac.jp}}

\vskip 0.4cm

{\it
$^1$ The Hakubi Center for Advanced Research and Graduate School of Informatics,\\ Kyoto University, Sakyo, Kyoto 606-8501, Japan\\
$^2$ Center for Complex Systems Biology, Universal Biology Institute, University of Tokyo, Meguro, Tokyo 153-8902, Japan
}

\vskip 1.5cm

\end{center}

\renewcommand{\thefootnote}{\arabic{footnote}}

\baselineskip 0.6cm


{\bf
If a cultural feature is transmitted over generations and exposed to stochastic selection when spreading in a population, its evolution may be governed by statistical laws and be partly predictable, as in the case of genetic evolution. Music exhibits steady changes of styles over time, with new characteristics developing from traditions. Recent studies have found trends in the evolution of music styles, but little is known about their relations to the evolution theory. Here we analyze Western classical music data and find statistical evolutionary laws. For example, distributions of the frequencies of some rare musical events (e.g.\ dissonant intervals) exhibit steady increase in the mean and standard deviation as well as constancy of their ratio. We then study an evolutionary model where creators learn their data-generation models from past data and generate new data that will be socially selected by evaluators according to the content dissimilarity (novelty) and style conformity (typicality) with respect to the past data. The model reproduces the observed statistical laws and can make non-trivial predictions for the evolution of independent musical features. In addition, the same model with different parameterization can predict the evolution of Japanese enka music, which is developed in a different society and has a qualitatively different tendency of evolution. Our results suggest that the evolution of musical styles can partly be explained and predicted by the evolutionary model incorporating statistical learning, which can be important for other cultures and future music technologies.
}

\section*{Introduction}

A prominent feature of humans is that they learn and transmit cultural traits over generations \cite{MajorTransitionsInEvolution}.
Although many cultural traits (e.g.\ style of music/language/fine art, fashion, unscientific beliefs, etc.) seem to make little direct contribution to an individual's biological fitness, some of them (e.g.\ music and fashion) have evolved into highly complex forms and have rather large influence on human behaviour.
To understand human's behaviour, it is important to uncover some possible laws in cultural evolution and seek for a theory that can explain them \cite{CultureAndEvolution,CulturalTransmissionEvolution}.
Moreover, a theory that can quantitatively predict parts of cultural evolution can serve as a base for the development of new technologies to predict cultural trend and enhance cultural production.

In this study, we consider the evolution of musical styles, which has gathered growing attention \cite{Serra2012,Zivic2013,Mauch2015,Honing2015,Ravignani2016,LeBomin2016,Savage2019}.
It has been observed in a recent paper \cite{Weiss2018} that some features of music, e.g.\ the frequency of tritones, have steadily increased during the history of Western classical music.
(The tritone is a pitch interval consisting of six semitones. It is regarded as a ``dissonant'' interval in traditional music theories \cite{HarmonyAndCounterpoint}.)
Although these clear trends imply some driving force for the evolution of the music style, their theoretical origins and how much they can be predicted are not understood.
Moreover, while previous studies \cite{Serra2012,Zivic2013,Mauch2015,Weiss2018} have focused on the mean of musical features, other statistics including the standard deviation and distribution form are important for studying their dynamics in relation to evolutionary models \cite{Lassig2017}.

\section*{Data Analysis}
\begin{figure}[t]
\centering
{\includegraphics[clip,width=0.95\columnwidth]{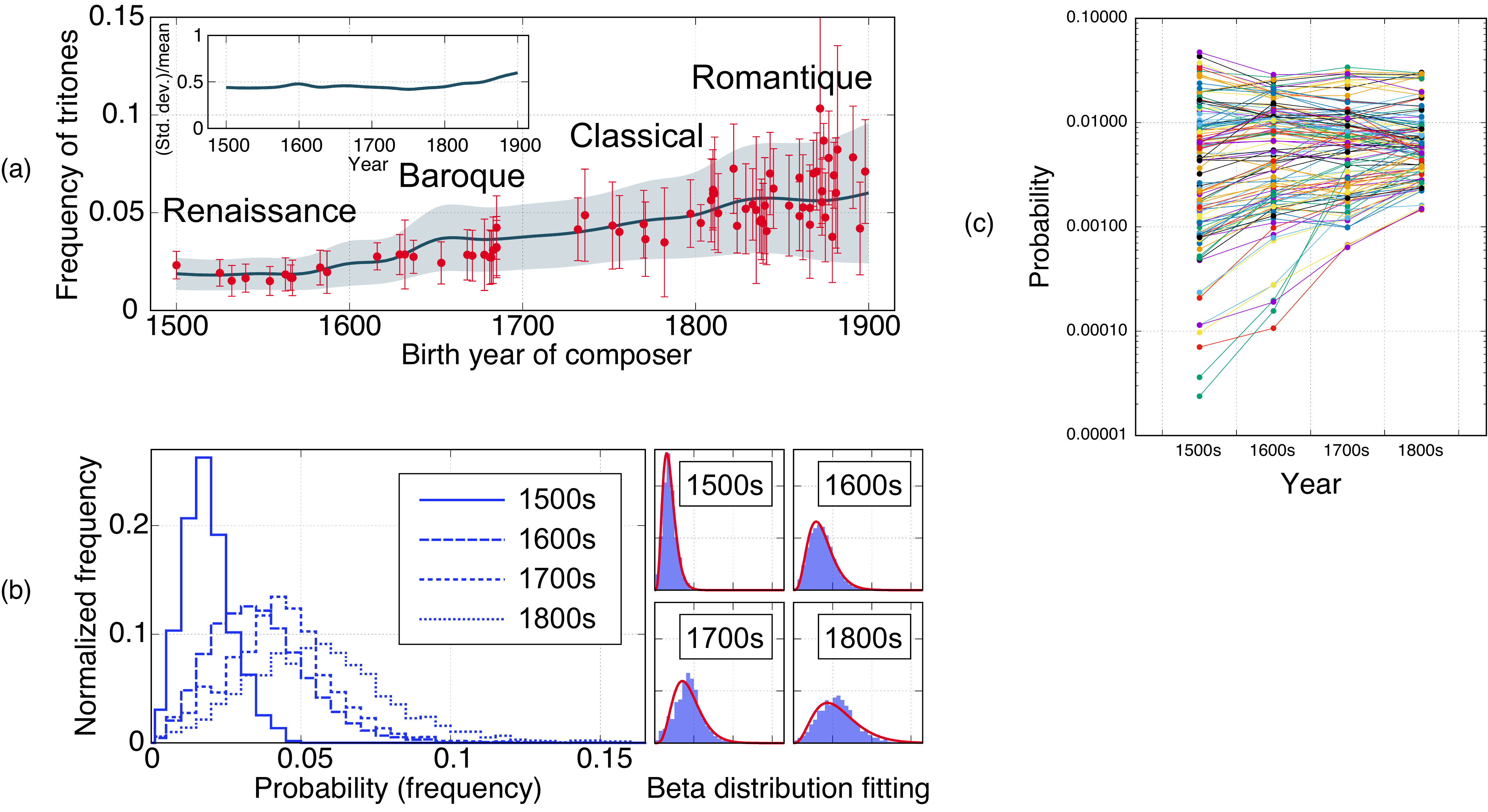}}
\vspace{-2mm}
\caption{(a)(b) Evolution of the distribution of frequencies of tritones observed in Western classical music data. In (a), points and bars indicate the mean and standard deviation for each composer, and the step line and shade indicate those for time windows of 100-year width shifted in units of 25 years (spline interpolation applied). In (b), distributions obtained for each century are shown with their best fit beta distributions. (c) Evolution of the means of frequencies of bigrams of pitch-class intervals.}
\label{fig:IntroData}
\end{figure}
We first analyze the time evolution of statistics of musical features of Western classical music data.
In particular, we study two major aspects of tonal content of music, dissonance and tonality, on which musicologists have focused \cite{TonalHarmony,HistoryOfWesternMusic}.
More specifically, we focus on the frequencies of two features that represent these aspects and can be computationally analyzed without interpretation of data by human.
One of them is tritone, which is a representative interval historically considered dissonant \cite{TonalHarmony} and has been studied also in a previous study \cite{Weiss2018} (see Refs.~\cite{Krumhansl1990,McDermott2016} for psychological studies on consonance and dissonance of musical sound).
The other one is non-diatonic motion, which is defined as bigrams of pitch-class intervals that cannot be realized on a diatonic scale.
(The C-major scale or ``the scale of white keys'' (C-D-E-F-G-A-B) is one instance of diatonic scales. In general, a diatonic scale can be transposed to the C-major scale by a global pitch shift.)
The non-diatonic motion is an indicator of chromatic motions and modulations (key changes) to a distant key \cite{Nakamura2015} (see Methods for a detailed definition).

Fig.~\ref{fig:IntroData}(a) shows that the mean and standard deviation of the frequency (probability) of tritones steadily increased during the years 1500--1900 while their ratio stayed approximately constant over that time (see Methods for details of the analysis).
Here and in what follows, zero-frequency data points are excluded from the analysis in order to obtain statistically reliable results.
Fig.~\ref{fig:IntroData}(b) shows actual distributions of the frequency of tritones obtained for each century.
We find that the distribution can be approximately fitted by a beta distribution (its function form is given in Eq.~(\ref{eq:CreatorModel})), which is a mathematically simple distribution defined over the range of real numbers between $0$ and $1$.

These statistical tendencies can also be found for the frequency of non-diatonic motions.
Statistical data of their frequency corresponding to Fig.~\ref{fig:IntroData} are given in Supplemental Material.
As a general set of musical features including these two, we can consider bigram probabilities of pitch-class intervals \cite{Nakamura2015}, which have 121 elements (see Methods).
Fig.~\ref{fig:IntroData}(c) shows how the means of these 121 features evolved over the centuries.
We see that low-probability features exhibit exponential-like growth.
Still, the number of observations is small and statistics of such rare events may not be so reliable.
(The distributions of very rare events typically have a peak at zero and the standard deviation is larger than the mean. This limitation of observation is caused by the fact that a musical piece usually consists of $10^2$--$10^3$ musical notes.)

To summarize our data analysis results, we have found the following statistical evolutionary laws in low-probability features of Western classical music data:
\begin{enumerate}\setlength\itemsep{-0.2em}
\item Beta-like distribution of frequency features
\item Steady increase of the mean and standard deviation
\item Nearly constant ratio of the mean and standard deviation, which is slightly less than unity
\item (Possibly) exponential-like growth of the mean
\end{enumerate}
These findings reveal that the evolution of styles in the classical music data exhibits much more regularities than previously found \cite{Weiss2018}.
The last two laws indicate that the dynamics is scale invariant, that is, the dynamics at one value of the features looks similar to that at a different value of the features.
Since these statistical laws are found in the music data of various composers in four consecutive centuries, they may be caused by general mechanisms of transmission and selection of cultural style rather than by the circumstances of individual composers or social communities of individual time periods.

\section*{Theoretical Model}

Let us now discuss a possible evolutionary model that may explain the origin of the observed statistical laws.
Following the general framework of Darwinian evolution, we construct a theoretical model based on information transmission and stochastic selection.
A feature of music culture is that creation styles are learned and transmitted via data (e.g.\ musical scores and audio signals), and recent studies have suggested the importance of statistical learning for music composition (e.g.\ \cite{Ebcioglu1990,Pachet2011,Tsushima2018}) and for lister's understanding (e.g.\ \cite{Ettlinger2011}).
As is commonly done in the field of music informatics (e.g.\ \cite{Pachet2011,Tsushima2018}), we represent creators (composers) with statistical models for data generation and try to capture the evolution of music styles through dynamic changes of creators' models.
As a driving force for time evolution, we consider social selection by contemporary evaluators (listeners).
Specifically, we study a dynamical system of creators that statistically learn their data-generation models from past data and then generate new data, and of evaluators that determine the fitness of the generated data.
Since evaluators should also learn their data-evaluation models from existing data, it is legitimate to consider a dynamic change in the fitness depending on other agents/data, as in evolutionary game theory \cite{EvolutionaryGames}.
Similar models of iterated learning have been studied in the context of language evolution \cite{Hashimoto1996,Kirby2001,Nowak2001}.

A dynamical system we call a {\it statistical creator-evaluator (SCE) model} is formulated as follows.
Each creator at generation $t$ generates a dataset $X_t$ of musical pieces according to a distribution (data-generation model) $\phi_t$, and the generated data are evaluated with the fitness defined below.
Following this evaluation, the creator's model of the next generation $\phi_{t+1}$ is determined by statistical learning.
With this procedure, the creator's data-generation model evolves over generations.
The creator's model $\phi_t(\theta)$ is defined over a probability parameter $\theta\in(0,1)$ (e.g.\ frequency of some musical events).
An evaluator is similarly modelled by a distribution $\psi_t(\theta)$.
We assume that $\phi_t$ is described as a beta distribution ($a_t,b_t>0$):
\begin{equation}
\phi_t(\theta)={\rm Beta}(\theta;a_t,b_t)\equiv\frac{1}{B(a_t,b_t)}\theta^{a_t-1}(1-\theta)^{b_t-1},
\label{eq:CreatorModel}
\end{equation}
The beta distribution is introduced here because it is a simple distribution function for the probability variable whose value is restricted between $0$ and $1$, where the emergence of a new feature is represented by a shift of a peak at $0$ towards $1$, and because it approximates the data well as shown in Fig.~\ref{fig:IntroData}(b).
For $a_t,b_t>1$, a beta distribution satisfies boundary conditions $\phi_t(0)=\phi_t(1)=0$ and the parameters $a_t$ and $b_t$ specify the power of $\theta$ at the boundaries.
The data-generation process is described in two steps: (i) a value of $\theta$ is drawn from $\phi_t(\theta)$ for each data unit (e.g.\ musical piece), here called a {\it (data) chunk}, and (ii) data samples (e.g.\ musical notes) in that chunk are sampled with the chosen $\theta$.
It is assumed that the data selection is carried out in the space of $\theta$ so that the model is described in terms of the space of $\theta$ without referring to data samples, so we treat $\theta$ as a directly observable quantity.

\begin{figure}[t]
\centering
{\includegraphics[clip,width=0.6\columnwidth]{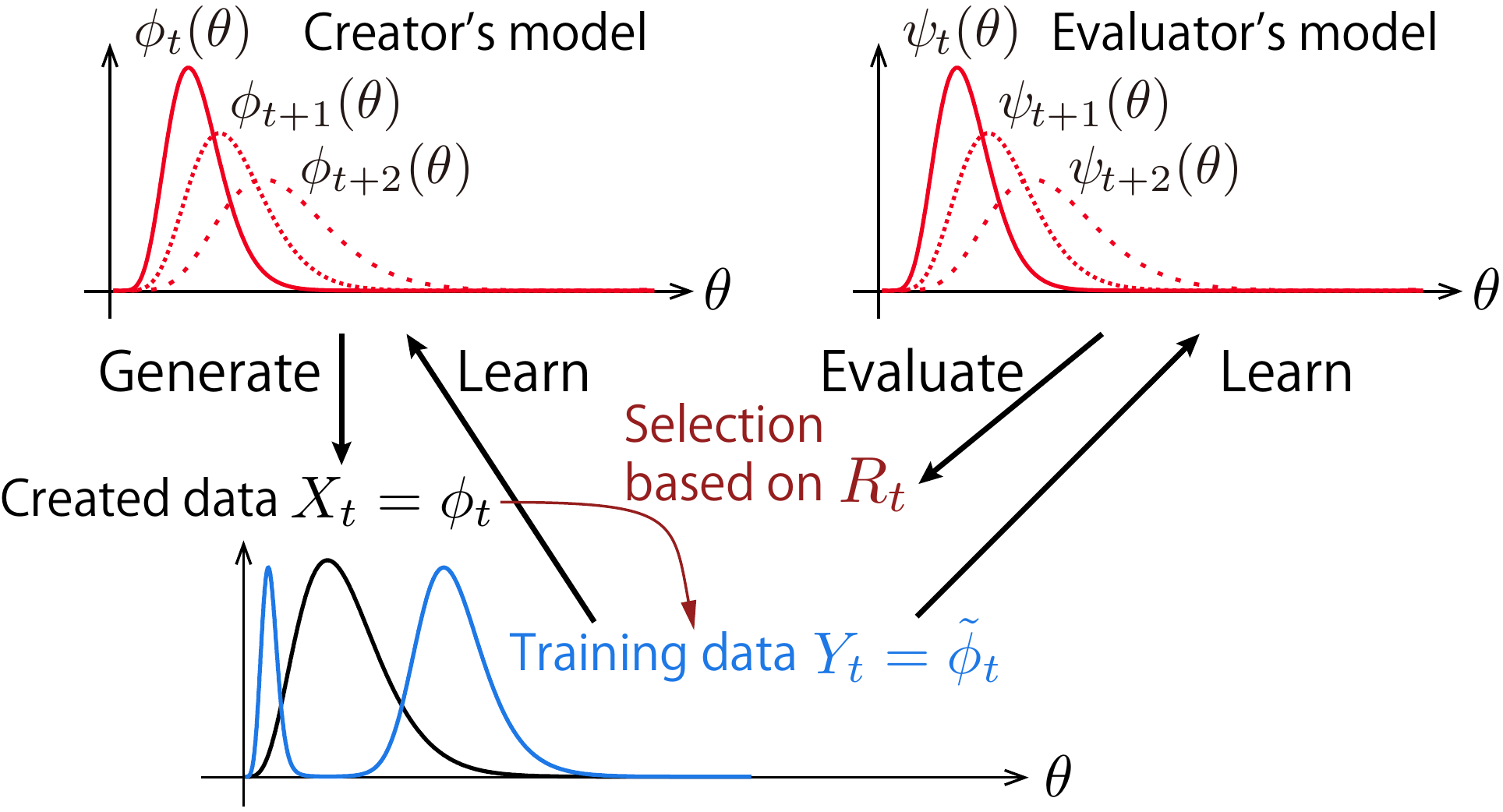}}
\vspace{-2mm}
\caption{Coevolution of creator's and evaluator's models through a social selection process modelled by the statistical creator-evaluator (SCE) model.}
\label{fig:SCEM}
\end{figure}
The data-selection (evaluation) process is described as follows (see Fig.~\ref{fig:SCEM}).
At each generation $t$, a dataset $X_t$ of chunks is generated by $\phi_t(\theta)$.
Data chunks in $X_t$ are then evaluated by the evaluators assigning weights $e^{\beta R_t(\theta)}$, where $R_t(\theta)$ is a functional of $\psi_t$ and $X_t$ called an evaluation function and $\beta$ is a selection coefficient.
The next-generation creator's model $\phi_{t+1}(\theta)$ and evaluator's model $\psi_{t+1}(\theta)$ are learned from the dataset of chunks denoted by $Y_t$ (i.e.\ $Y_t$ is used as training data).
The data $Y_t$ consists of data chunks selected randomly from $X_t$ with probabilities proportional to $e^{\beta R_t(\theta)}$.
It is assumed that $\phi_t$ and $\phi_{t+1}$ have the same distribution form (\ref{eq:CreatorModel}) and the parameters of $\phi_{t+1}$ are chosen to approximate $Y_t$ as much as possible (the learning scheme is specified later), even though the data $Y_t$ are distributed differently from $\phi_{t+1}$ in general.
In the limit of infinite data size, $X_t$ is distributed as $\phi_t(\theta)$ and $Y_t$ is distributed in proportion to $\phi_t(\theta)e^{\beta R_t(\theta)}$.
Here, we consider a simple case where $\psi_t(\theta)$ is learned in the same way as $\phi_t(\theta)$ so that these distributions are in fact identical.
The dynamics is summarized as
\begin{equation}
\phi_{t+1}(\theta)=\psi_{t+1}(\theta)\leftarrow \tilde{\phi}_t(\theta):=\phi_t(\theta)e^{\beta R_t(\theta)}.
\label{eq:Update}
\end{equation}
where the arrow means that the distribution on the left-hand side is learned from the data on the right-hand side.
Although we mainly focus on the case where $\phi_t$ is given as in Eq.~(\ref{eq:CreatorModel}), $\phi_t$ in the SCE model in Eq.~(\ref{eq:Update}) can be described with other distributions in general.

Since the fundamental process of evaluating musical data is unknown, we attempt to derive a reasonable form of the evaluation function $R_t$ based on a theoretical argument.
Rather than introducing biases depending on particular features of music content, we here focus on two viewpoints for evaluation considered most fundamental and general \cite{CultureAndEvolution}: the content dissimilarity (novelty) and style conformity (typicality) with respect to the past data.
Naively, novelty is important because newly generated data chunks whose content is very similar to that of an existing data chunk do not increase the experience of the evaluators and thus are not favoured.
The fast updates of popular music album charts suggest the possibility of this bias \cite{Bentley2007}.
Typicality is also important because a data chunk that deviates significantly from the style of the past data cannot be understood easily by the evaluators and are not approved.
Some critics' denials of innovative musical works such as Berlioz's Symphonie Fantastique \cite{Berlioz} and Stravinsky's Rite of Spring \cite{HistoryOfWesternMusic} at their premieres suggest the relevance of this bias.

We propose to mathematically formulate these two metrics in terms of information measures.
Novelty can be formulated by considering the effective amount of information obtained from the evaluator's perspective.
For each value of $\theta$, novelty can be measured with the amount of similar data chunks in $X_t=\phi_t$, which is proportional to $\phi_t(\theta)$ in the limit of infinitesimal precision of discriminating musical features (see Methods for a detailed derivation).
Typicality can be formulated by considering the difficulty of understanding, or memorizing, a data chunk according to the evaluator's model $\psi_t$.
Thus, in information-theoretical terms, typicality can be described as the number of bits needed to encode the information contained in a data chunk $\theta$ using the model $\psi_t$, which is proportional to $-{\rm ln}\,\psi_t(\theta)$ \cite{CoverThomas}.

In other words, to gather the information contained in a data chunk $\theta$, the evaluator must first spend cost proportional to $\phi_t(\theta)$ to obtain that data chunk (together with unavoidable similar data chunks) and then spend cost proportional to $-{\rm ln}\,\psi_t(\theta)$ to memorize the contained information.
In this way, the evaluation function constructed as a sum of the novelty and typicality defined here can be interpreted as the effective amount of cost necessary for the evaluator to gather information.
In this sense, the novelty and typicality biases may have relation to biological fitness, as the ability to gather information about the environment is essential for surviving.

By using the analogy of the above selection probability with a Boltzmann distribution in statistical physics, where $\beta$ and $R_t$ correspond to the inverse temperature and negative energy (cost), the form of $R_t$ is given as
\begin{equation}
\beta R_t(\theta)=\beta_T\,{\rm ln}\,\phi_t(\theta)-\beta_N\phi_t(\theta),
\label{eq:EvaluationFunction}
\end{equation}
where $\beta_T$ and $\beta_N$ are constant factors, the first and second terms respectively represent the typicality and novelty of chunk $\theta$, and we have used the relation $\psi_t=\phi_t$.
Substituting Eq.~(\ref{eq:EvaluationFunction}) into Eq.~(\ref{eq:Update}), we have
\begin{equation}
\tilde{\phi}_t(\theta)=\phi_t(\theta)^{1+\beta_T}{\rm exp}[-\beta_N\phi_t(\theta)].
\label{eq:OneDimLGS}
\end{equation}
The signs of the two terms in Eq.~(\ref{eq:EvaluationFunction}) are chosen so that when $\beta_T$ and $\beta_N$ are positive, evaluators favour both typical and novel data chunks.
Theoretically, these parameters can take negative values in general.

To complete a mathematical formulation, we specify the learning process.
The creator learns the data distribution $\phi_{t+1}(\theta)$ from $\tilde{\phi}_t(\theta)$ so that $\phi_{t+1}(\theta)$ is assimilated by the beta distribution by optimizing its parameters.
To be specific, noting that the pair $(a_t,b_t)$ has one-to-one correspondence with the pair of mean and standard deviation $(\mu_t,\sigma_t)$ (see Methods), we use the moment matching method to learn $\phi_{t+1}$ from $\tilde{\phi}_t$.
That is, we choose the parameters of $\phi_{t+1}$ so that its mean $\mu_{t+1}$ and standard deviation $\sigma_{t+1}$ exactly match those of $\tilde{\phi}_t$.
If we take the statistics $\mu_t$ and $\sigma_t$ as state variables, the update equation (\ref{eq:Update}) is described as a two-dimensional map $(\mu_t,\sigma_t)\to(\mu_{t+1},\sigma_{t+1})$.

Let us analyze the model. See Methods for mathematical details.
Qualitatively, positive $\beta_T$ and negative $\beta_N$ put higher weights on more probable $\theta$, causing $\phi_{t+1}$ to be sparser than $\phi_t$.
Conversely, negative $\beta_T$ and positive $\beta_N$ make $\phi_{t+1}$ less sparse.
The case $\beta_T<-1$ puts infinite weights on zero-probability $\theta$ and is thus ill-defined.
In the following, we focus on the case $\beta_T,\beta_N\geq0$, $\mu_t<1/2$, and $\sigma_t<\mu_t$, and in particular the regime where $\mu_t$ is small, to analyze the dynamical system quantitatively.

\begin{figure}[t]
\centering
{\includegraphics[clip,width=0.93\columnwidth]{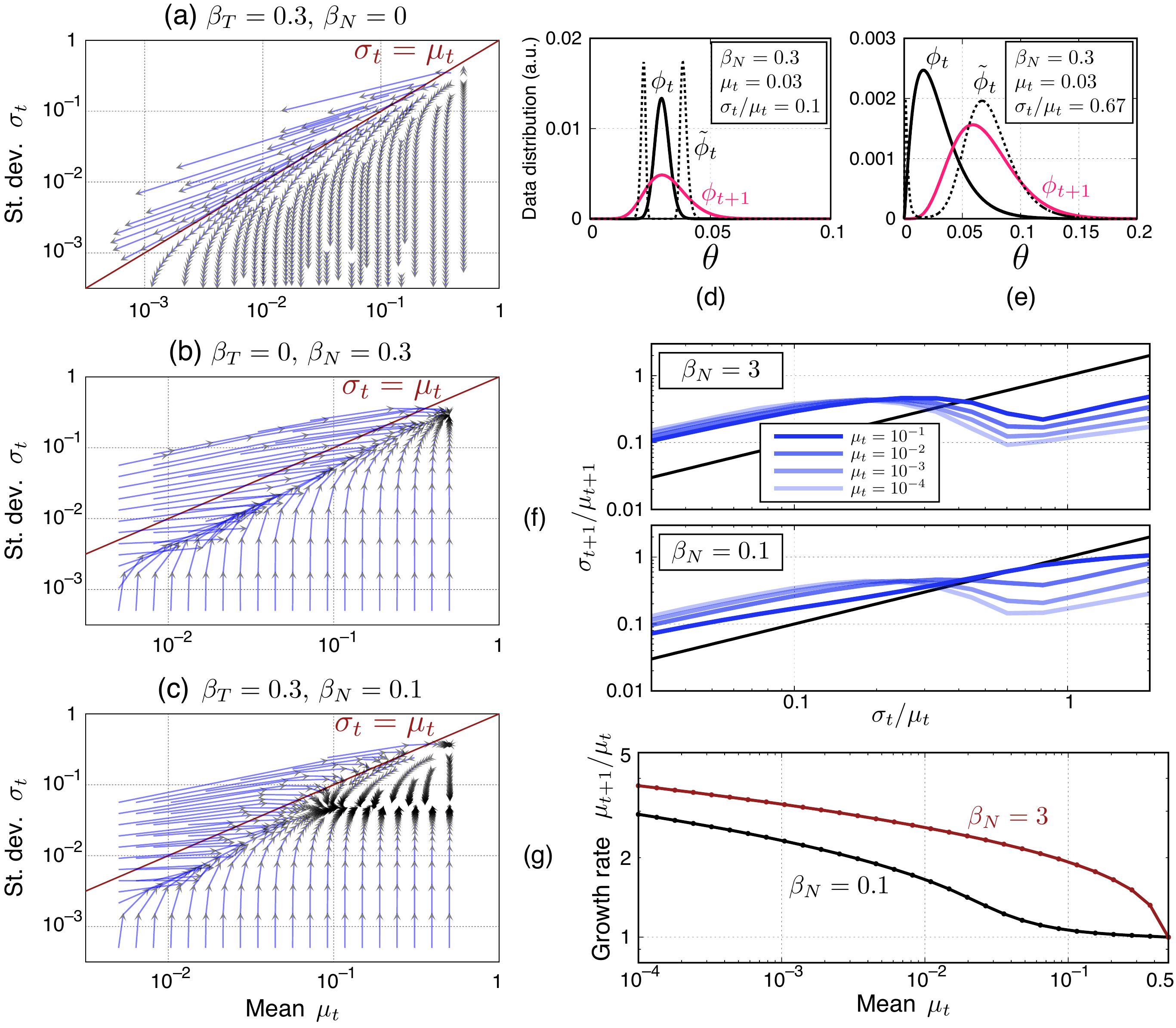}}
\vspace{-2mm}
\caption{(a)--(c) Orbits of the SCE model for three cases of $\beta_T$ and $\beta_N$.
(d)(e) Examples of an update in the case $\beta_N>0$ and $\beta_T=0$ for small and large $\sigma_t/\mu_t$.
(f) Dynamics of the ratio $\sigma_t/\mu_t$.
(g) Growth of the mean around the slow manifold ($\sigma_t/\mu_t=0.6$).
}
\label{fig:2}
\end{figure}
For small $\beta_T$ and $\beta_N$, which are of our interest, the discrete-time dynamics of the system is relatively smooth and vectors in Fig.~\ref{fig:2}(a)--(c) show how an update changes $\mu$ and $\sigma$ at each point.
When $\beta_N=0$ (i.e.\ only typicality is evaluated), both the mean and standard deviation decrease over time (Fig.~\ref{fig:2}(a)).
More specifically, the mean will converge to the mode (peak position) whereas the standard deviation will converge to $0$ for $t\to\infty$.
This is shown analytically in Methods.

When $\beta_N>0$ and $\beta_T=0$ (i.e.\ only novelty is evaluated), both the mean and standard deviation increase over time and the orbits converge to a fixed point with $\mu_{t=\infty}=1/2$ (Fig.~\ref{fig:2}(b)).
The reason the mean increases can be understood intuitively from the shape of the distribution.
When $\beta_N$ is not too small, the weighted data $\tilde{\phi}_t$ has two peaks around the mean of $\phi_t$ and the left one is narrower due to the boundary at zero (as in Fig.~\ref{fig:SCEM}) so that the distribution $\phi_{t+1}$ is pushed to the right.

A notable feature of this case is the presence of a ``slow manifold''.
The dynamics quickly fall onto the manifold (i.e.\ subspace of the parameter space) with $\sigma_t/\mu_t\approx{\rm constant}$, which is slightly less than unity.
The values of $\mu_t$ and $\sigma_t$ will then grow along the manifold keeping their ratio almost constant in time.
Intuitively speaking, this slow manifold is formed because when $\sigma_t\ll\mu_t$ the beta distribution is almost symmetric and an update does not change $\mu_t$ significantly but increases $\sigma_t$ and thus also $\sigma_t/\mu_t$ (Fig.~\ref{fig:2}(d)).
When $\sigma_t\sim\mu_t$, the right peak of $\tilde{\phi}_t$ dominantly influences the next distribution $\phi_{t+1}$ and $\mu_t$ grows so much that $\sigma_t/\mu_t$ decreases (see Fig.~\ref{fig:2}(e) and Methods for more details).
This is quantitatively shown in Fig.~\ref{fig:2}(f), where one can see that the curve representing the update of the ratio $\sigma_t/\mu_t$ intersects with the invariant line at similar points for varying $\mu_t$.
As can be observed in Fig.~\ref{fig:2}(f) and will be discussed more analytically in Methods, the constant value of $\sigma_t/\mu_t$ is smaller for larger $\beta_N$.

How the mean grows on the slow manifold can be understood from Fig.~\ref{fig:2}(g).
One finds that, for various values of the mean, its growth rate is of the same order of magnitude.
This indicates that the mean (and thus also the standard deviation) grows nearly exponentially over time.
The comparison between different values of $\beta_N$ in Fig.~\ref{fig:2}(g) shows that the growth rate is not very sensitive to the value of $\beta_N$.

The model's dynamics for finite $\beta_T$ and $\beta_N$---i.e.\ when both novelty and typicality are evaluated---are illustrated in Fig.~\ref{fig:2}(c).
Generally, the standard deviation converges to a fixed point where the effects of the typicality and novelty terms balance; when the standard deviation is larger (smaller) than its asymptotic value the dynamics is similar to that of the typicality (novelty) term only.
In particular, for small $\mu_t$ and $\sigma_t$ (i.e.\ in the early stage of evolution), we again find a slow manifold where both the mean and standard deviation eventually increase while their ratio stays almost constant.
If $\sigma_t\ll\mu_t$ when $\sigma_t$ reaches the fixed point, then the value of $\mu_t$ is effectively frozen, leading to the emergence of marginally stable points.

One can also confirm the presence of a similar slow manifold in the case of $\beta_N>0$ for other choices of $\phi_t$ with a boundary at $\theta=0$, i.e.\ the gamma and log-normal distributions (see Supplemental Material).
This shows that it is a rather general phenomenon, as expected from the aforementioned intuitive argument.

\section*{Examining the Model with Experimental Data}

Let us now compare the consequences of the present model with the observed data of music evolution.
Among the aforementioned four statistical evolutionary laws, the first law (beta distribution) is naturally incorporated in the model.
When the novelty term is present and the initial values satisfy $\sigma<\mu\ll1$, the dynamics of the model spontaneously lead to the phase where both the mean and standard deviation increase over time keeping their ratio almost constant and slightly less than unity (the second and third laws).
This explains the origin of these laws, which are expected by the model {\it irrespective} of small changes in initial values.
We have also shown that the last law (exponential growth) is also derived from the dynamics of the model in the early stage of evolution.

To illustrate the characteristics of the present model and to examine the model's predictive ability, let us briefly discuss another evolutionary model, for which the evaluation function $R_t$ is simply a function of $\theta$, rather than a functional of $\psi_t$ or $\phi_t$ as in the SCE model.
Since the constancy of $\sigma/\mu$ suggests scale-invariant dynamics, we consider an evaluation function with a log potential: $R_t={\rm ln}\,\theta$.
As a natural choice for $\phi_t$, we here use the log-normal distribution instead of the beta distribution because it is kept invariant under the selection process and its shape is similar to that of the beta distribution (see Supplemental Material for a graphical comparison between these distributions).
The creator's model is then written as
\begin{equation}
\phi_t(\theta)=\frac{1}{\sqrt{2\pi}\tilde{\sigma}_t\theta}{\rm exp}\bigg[-\frac{({\rm ln}\,\theta-\tilde{\mu}_t)^2}{2\tilde{\sigma}^2_t}\bigg],
\label{eq:LogNormal}
\end{equation}
where $\tilde{\mu}_t$ is the log mean and $\tilde{\sigma}^2_t$ is the log variance, which are related to the mean and standard deviation as $\mu_t={\rm exp}(\tilde{\mu}_t+\tilde{\sigma}^2_t/2)$ and $\sigma_t/\mu_t=\sqrt{{\rm exp}(\tilde{\sigma}^2_t)-1}$.
We call this model the {\it log-potential model}.

As shown in Methods, in this model the standard deviation and mean both grow exponentially over time with a fixed rate $e^{\beta\tilde{\sigma}^2}$ and their ratio $\sigma_t/\mu_t$ is kept constant.
While the dynamics of the log-potential model is similar to that of the SCE model on the slow manifold, an important difference is the sensitivity to the initial values and selection coefficient.
For the log-potential model the ratio $\sigma_t/\mu_t$ can be tuned arbitrarily by adjusting the initial condition, whereas for the SCE model it is driven to an asymptotic value spontaneously by the dynamics irrespective of the initial condition.
The growth rate $e^{\beta\tilde{\sigma}^2}$ is very sensitive to the value of $\beta$ in the log-potential model, whereas it is not very sensitive in the SCE model as we discussed above.

\begin{figure}[t]
\centering
{\includegraphics[clip,width=0.95\columnwidth]{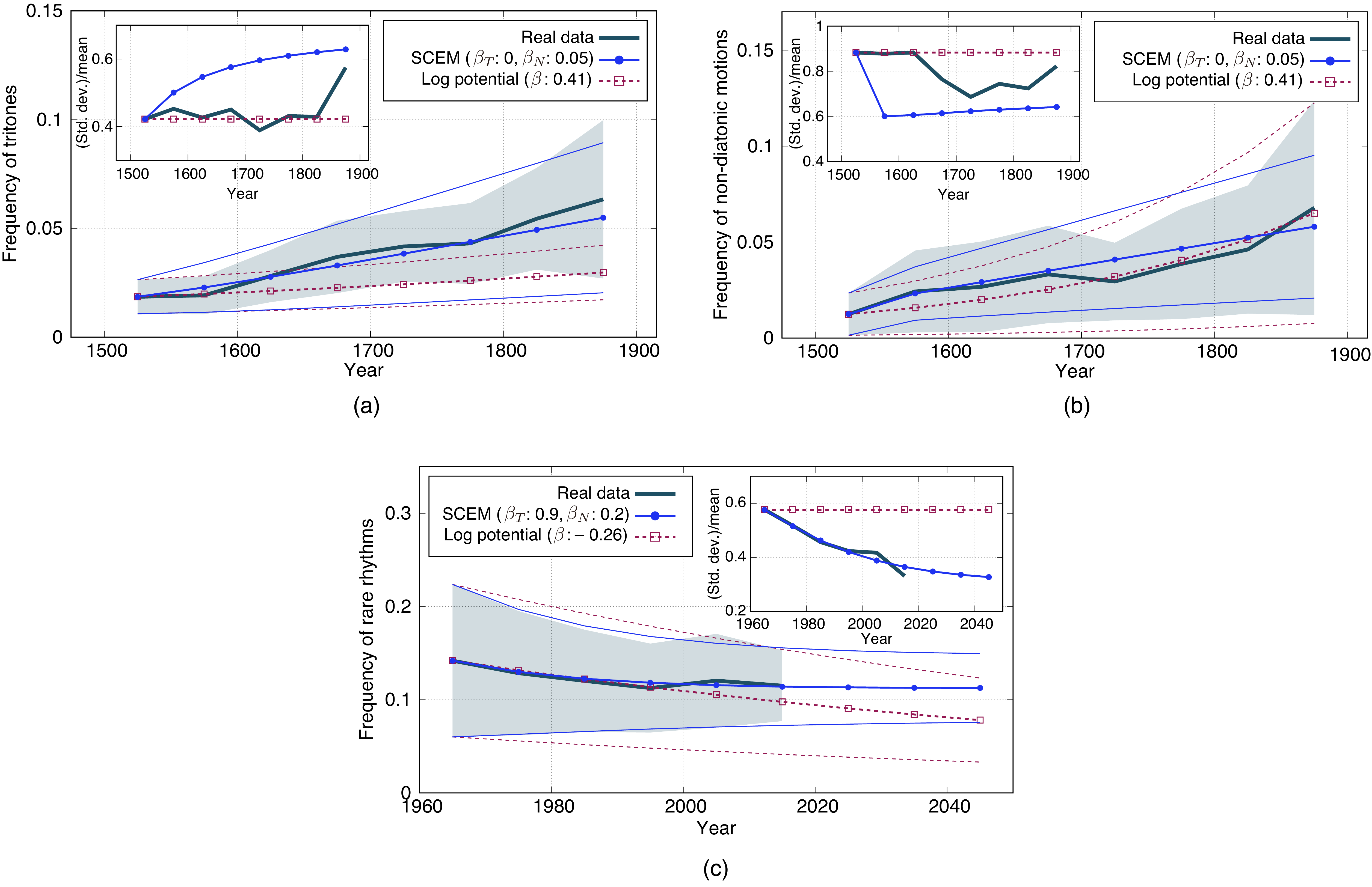}}
\vspace{-2mm}
\caption{Comparisons between model predictions and real data. Bold lines indicate means, and thin lines (the SCE model (SCEM) and log-potential model) and shadow (real data) indicate the ranges of $\pm1$ standard deviation. Model parameters are optimized to minimize the squared error of predicted means and standard deviations (optimal parameters are shown in the insets).
In (a) and (b), the model parameters are optimized simultaneously to fit the two datasets (the frequency of tritones and that of non-diatonic motions for the classical music data).
In (c), the time evolution of the frequency of rare rhythms in the enka music data is analyzed and compared with the models' predictions.
}
\label{fig:3}
\end{figure}
In Figs.~\ref{fig:3}(a) and \ref{fig:3}(b), the time evolution of the frequency of tritones and that of non-diatonic motions in the classical music data is numerically compared with the solutions of the SCE model and the log-potential model.
The models are initialized with the mean and standard deviation at the earliest time and the model parameters ($\beta_T$ and $\beta_N$ for the SCE model, and $\beta$ for the log-potential model) are optimized to minimize the squared error of the means and standard deviations throughout the time period of the data.
If the evolutions of these two features share the same mechanism, it is reasonable to use the same model parameters to fit both sets of data.
The parameters are thus optimized to fit both sets of data simultaneously and the optimized values are given inside each figure.

We see that the SCE model can roughly fit both sets of data whereas the log-potential model can fit only one set of data.
Quantitatively, the root mean squared errors for the tritone and non-diatonic motion data are $4.5\times10^{-3}$ and $1.5\times10^{-2}$ for the SCE model, and $7.2\times10^{-3}$ and $6.1\times10^{-3}$ for the log-potential model, respectively.
This result indicates not only that the two sets of data can be explained/predicted by the mechanism described by the SCE model in a unified way but also that the prediction is not trivial.
On the other hand, we also see some discrepancies between data and model predictions (e.g.\ in the values of $\sigma/\mu$).
Such small discrepancies can be explained in several ways: removing the simplifications assumed for the SCE model may bring small changes in model predictions, as discussed later, and they may be simply due to statistical/systematic error in the data.
If we try to fit the two sets of data individually using different parameter values, the fitting error for the log-potential model is slightly smaller than that for the SCE model (see Supplemental Material).

To examine the ability of the model with different data, Fig.~\ref{fig:3}(c) illustrates results of another analysis on a different musical feature extracted from a different dataset.
The dataset is a collection of enka music (a genre of Japanese popular music) compiled and published by a music publisher \cite{EnkaMale,EnkaFemale}.
Here we focus on the rhythms and use as a feature the frequency of ``rare rhythms'' that are defined as bigrams of note values whose ratio is not one of $\{1,1/2,2,2/3,3/2,1/3,3,1/4,4,1/6,6\}$ (see Methods for details).
Both the mean and standard deviation decrease over time, which is qualitatively different from the previous two cases.
For this case, only the SCE model can reproduce the history of the mean and standard deviation.
Predictions for the near future are also provided in Fig.~\ref{fig:3}(c).
As expected from the decrease of the standard deviation, typicality plays a more influential role for data selection and the model makes a testable prediction that the mean will converge in the future.
One interpretation of this result is that enka music is considered as a kind of ``soul music'' \cite{EnkaBook} and the evaluators (listeners) would prefer a typical enka song over a novel one.
On the other hand, the log-potential model predicts a linear-like decrease of the mean, which can be discriminated from the prediction of the SCE model in a near future.
The results show the nontrivial ability of the SCE model to explain and predict the nonlinear evolution of the enka music and demonstrate that the model is relevant not only for Western classical music but also for music developed in a different cultural background.

\section*{Discussion}

In conclusion, we have analyzed Western classical music data and found several statistical evolutionary laws, in particular, steady increase of the mean and standard deviation of frequencies of rare events and nearly constancy of their ratio, which indicate some driving force for the evolution of the music style.
As a theoretical explanation of the phenomenon, we have formulated and analyzed SCE models in which creators and evaluators coevolve by influencing each other through a social selection process.
The evaluation function for the social selection is constructed with the novelty and typicality terms representing cost required for obtaining and memorizing data in the process of information gathering.
We have shown that when the creator's and evaluator's models are beta distributed as observed in real data and the novelty term is active, the system generally has a slow manifold in which both the mean and standard deviation grow almost exponentially while their ratio stays almost constant.
This property and the fact that the system's dynamics are relatively insensitive to the selection coefficients make the present model more predictive and distinct from a Darwinian evolutionary model with a logarithmic potential.

It has been demonstrated that the present model can predict the evolution of the Western classical music data better than a scale-invariant evolutionary model (log-potential model).
The present model had the ability to fit the two kinds of data (frequency of tritones and that of non-diatonic motions) in a unified way, whereas the log-potential model could only fit the data individually.
From the perspective of the present model, the observed evolution of the mean and standard deviation of frequencies of musical features that were once rare is a consequence of pursuing novelty.
For the dataset of enka music, both the mean and standard deviation of the frequency of rare rhythms were found to be decreasing, which indicated that typicality has more importance than novelty in the selection process.
Predictions for the evolution of this feature that are testable in the next few decades have been made.

In the evolutionary process studied here, the balance between novelty and typicality (i.e.\ content dissimilarity and style conformity with respect to past data) plays an essential role.
As we saw in the classical music data and enka music data, relative values of $\beta_N$ and $\beta_T$ can influence the direction and speed of evolution.
We also found that the ratio $\sigma/\mu$ of the standard deviation and mean is an important metric of evolutionary dynamics, which can be used to infer from data the relative importance of novelty and typicality in the process of social selection/evaluation.
Once these parameters are determined, the SCE model can be used to predict the evolution of musical features.
Such predictive ability opens the possibility of new technologies such as hit song prediction \cite{Herremans2014} and automatic composition systems that go beyond the ability of simply imitating the style of fixed training data \cite{Ebcioglu1990,Pachet2011} and generate next-generation music.

Since the novelty and typicality biases represent the effective amount of cost necessary for the evaluator to gather information and are not dependent on particular features of music, they can be important for other types of culture, and the present model can be useful for analyzing not only music data but also other cultural data.
Evolutionary dynamics of language \cite{Michel2011}, other genres of music \cite{Mauch2015}, scientific topics \cite{Griffiths2004}, and sociological phenomena \cite{Castellano2000,Abrams2011} are among topics currently under investigation.
Another relevant topic is the evolution of bird songs, where selection-based learning is considered important \cite{Nelson1994}.
Evolutionary dynamics of bird songs have been studied based on dynamical systems that describe interaction between generators (singing birds) and imitators \cite{Suzuki1994}, which is similar to the novelty-typicality bias in this study.

Several remarks are made before closing the paper.
First, there are multiple possible ways of extending and relaxing the condition of the minimal model analyzed in this study.
Relaxing the assumption that both the creator and evaluator learn from the same data can lead to time displacement of their models.
For example, if the evaluator learns its model $\psi_{t+1}$ directly from the data $X_t$, instead of being biased by the evaluation function, then $\psi_{t+1}=\phi_t$ holds.
We can also introduce overlaps between generations or dependence on data created by more than one past generation.
These extensions can change the consequences of the model quantitatively and can possibly explain the small discrepancies between model prediction and data in Fig.~\ref{fig:3}.
Systems with multiple creators and evaluators would also be important for investigating the diversification and specification of cultural styles.

Second, a way to test the present model is to observe the exponential growth of a relevant feature.
However, this is not easy for music data because of the size of each data chunk (musical piece) is relatively small.
A musical piece typically consists of $10^2$ to $10^3$ notes and thus observing the evolution of a frequency of musical events across some orders of magnitude is difficult due to data sparseness.
It would be possible to alleviate the problem by extending creator's and evaluator's models in a Bayesian manner.
Another direction for experimentally testing the model is to directly examine the evaluation function by means of music data with social rankings, by an evolutionary experiment involving humans as evaluators \cite{MacCallum2012}, or by psychological experiments \cite{Huron2006}.
It would also be possible to infer the form of the evaluation function from such data by machine-learning techniques.

Third, one might think that music styles are transmitted via a set of rules (often called music theories) and the SCE model does not accord with the reality.
It has been argued that traditional composition rules are not sufficient to describe the actual composition process from a computational viewpoint \cite{Ebcioglu1990}, and in fact traditional music theories tell little about the quantitative nature of music styles \cite{HistoryOfMusicalStyle,DeLaMotte}.
In addition, recent studies on music informatics have suggested that traditional composition rules can be acquired from data via statistical learning \cite{Allan2005,Tsushima2018}.
Based on these observations, our view is that although those composition rules may influence the transmission of music styles, the effect of statistical learning is essential for understanding the evolution of music styles.


Fourth, there are potential sources of social influence that could affect the evolution of musical styles other than the novelty and typicality biases studied here.
These sources include effect of random copying (neutral drift) in a finite population \cite{Bentley2007}, interdependence among evaluators' decision \cite{Salganik2006}, active role of creators \cite{Claidiere2012}, indirect biases (e.g.\ publicity) independent of the data content \cite{CultureAndEvolution}, and psychological biases related to specific musical features \cite{Herremans2014}.
Our results do not exclude the relevance of these sources to the studied phenomena of music evolution.
The contribution of this work is to propose another possible source of cultural evolution that can be particularly important when statistical learning is involved and to provide theoretical results that help identify its relevance in observed data.
Further research is necessary to study the consequences of the SCE model when it is extended to incorporate those other sources of mutation and social selection and to identify their individual roles in evolutionary dynamics.
It is also important to seek for a fundamental model that can explain an evolutionary origin of the novelty and typicality biases in Eq.~(\ref{eq:EvaluationFunction}) and can validate the assumption of the Boltzmann distribution in Eq.~(\ref{eq:Update}) as well as the beta distributions observed in the data.

\section*{Methods}
\label{sec:Methods}

\subsection*{Data Analysis}

A collection of Western classical musical pieces is used for the data analyses in Figs.~\ref{fig:IntroData} and \ref{fig:3}.
The dataset consists of MIDI files of 9,727 pieces by 76 composers downloaded from a public web page (\url{http://www.kunstderfuge.com}).
This dataset is compiled in order to cover a longer period of time than the datasets used in previous studies \cite{Zivic2013,Weiss2018} and to enable noiseless symbolic music analysis.
(The dataset used in \cite{Weiss2018} consists of audio data and the Peachnote corpus used in \cite{Zivic2013} contains symbolic music data that are obtained by scanned sheet music by using music optical character recognition (OCR) software and thus contain noise.)
The 76 composers are those with the largest number of available pieces and obvious duplications of two or more files for the same piece are avoided by looking at file names.
Files with less than 100 musical notes are also removed.
Each MIDI file is parsed and a sequence of pitches represented by integers in units of semitones is extracted; pitches are ordered according to their appearance in the file.
To extract information on music styles that are irrelevant of superficial features such as pitch range and absolute key, the sequence of pitch-class intervals is obtained.
Pitches are converted to pitch classes by applying a modulo operation of divisor $12$.
Then pitch-class intervals are obtained by taking the difference between adjacent pitch classes.
Note that these intervals include both melodic intervals and harmonic intervals, which are not distinguished in our analysis.
Finally, zero intervals, which correspond to successions of the same pitch or octave transitions, are dropped because they dilute other relevant features.
Each musical piece is now represented as a sequence of pitch-class intervals denoted by $\bm x=(x_1,\ldots,x_N)$.
Since zero intervals are excluded, there are $11$ types of unigrams and $121=11^2$ types of bigrams for pitch-class intervals.

In Figs.~\ref{fig:IntroData}(a), \ref{fig:IntroData}(b), and \ref{fig:3}(a), the frequency of tritones for each piece is defined as $\#\{n|x_n=6\}/|\bm x|$, where $|\bm x|$ denotes the number of elements in $\bm x$.
Although melodic tritones and harmonic tritones are not distinguished in our analysis, both their uses were once severely restricted in the medieval period and became increasingly common in later time periods \cite{HistoryOfWesternMusic,DeLaMotte}.
In Fig.~\ref{fig:3}(b), the frequency of non-diatonic motions is defined as $\#\{n|(x_n,x_{n+1})\in C\}/|\bm x|$, where the set of non-diatonic motions $C$ consists of the following 20 elements: $(1,1)$, $(1,3)$, $(1,8)$, $(1,10)$, $(2,11)$, $(3,1)$, $(3,8)$, $(4,4)$, $(4,9)$, $(4,11)$, $(8,1)$, $(8,3)$, $(8,8)$, $(9,4)$, $(9,11)$, $(10,1)$, $(11,2)$, $(11,4)$, $(11,9)$, and $(11,11)$.
It can be shown by direct calculation that these bigrams of pitch-class intervals represent non-diatonic motions that cannot be realized by note transitions on a diatonic scale \cite{Nakamura2015}.
Each of the 121 bigram probabilities in Fig.~\ref{fig:IntroData}(c) is similarly defined as the frequency of each possible pair $(x_n,x_{n+1})$.

For the result shown in Fig.~\ref{fig:3}(c), a dataset consisting of 761 songs of Japanese enka music is used \cite{EnkaMale,EnkaFemale}.
Each musical piece is first notated in the MusicXML format and then the sequence of note values (note lengths written in musical scores) is obtained.
The ratio between adjacent note values often has simple ratios such as $1$, $1/2$, $2$, $2/3$, and $3/2$.
Similarly as we look at rare pitch events like tritones and non-diatonic motions for the Western classical music data, we observe the frequency of rare rhythms, which are defined as bigrams of note values whose ratio is not one of $\{1,1/2,2,2/3,3/2,1/3,3,1/4,4,1/6,6\}$.

The statistics obtained from the two datasets used to create Figs.~1 and 4 are available at \url{https://evomusstyle.github.io/} .
Although the raw music data cannot be published due to the copyright issue for both datasets, the lists of pieces used for the analysis are provided, from which one can in principle reproduce the exact data we have.

\subsection*{Model Formulation}

In the discussion above Eq.~(\ref{eq:EvaluationFunction}), it is postulated that the novelty term is described as the amount of the cost of obtaining data chunks containing ``similar'' information in the generated data $X_t=\phi_t$.
To express this mathematically, we introduce a function $G(\theta,\theta')$ that measures the similarity between data chunks $\theta$ and $\theta'$.
Assuming that obtaining each data chunk requires the same amount of cost, the total cost of obtaining data chunks similar to $\theta$, denoted by ${\sf Novelty}(\theta)$, is given as
\begin{equation}
{\sf Novelty}(\theta)\propto\int d\theta'\,G(\theta,\theta')\phi_t(\theta')=\int d\theta'\,G(0,\theta'-\theta)\phi_t(\theta'),
\end{equation}
where we have assumed translational invariance in the last expression.
When evaluators can discriminate musical features with infinite precision, $G(0,\theta)$ is proportional to the delta function $\delta(\theta)$.
In this case, we have
\begin{equation}
{\sf Novelty}(\theta)\propto\phi_t(\theta)\int d\theta'\,G(0,\theta').
\end{equation}
Since the integral is constant with respect to $\theta$, we have shown that the novelty term is proportional to $\phi_t(\theta)$.

\subsection*{Model Analysis}

The parameters $a_t$ and $b_t$ of the beta distribution are in one-to-one correspondence with the mean $\mu_t$ and standard deviation $\sigma_t$ as follows.
\begin{equation}
\mu_t=\frac{a_t}{a_t+b_t}
\label{eq:muFromAB}
\end{equation}
\begin{equation}
\sigma_t=\frac{1}{a_t+b_t}\sqrt{\frac{a_tb_t}{a_t+b_t+1}}
\label{eq:sigmaFromAB}
\end{equation}
\begin{equation}
\sigma_t/\mu_t=\sqrt{(1-\mu_t)/(a_t+\mu_t)}
\end{equation}
We focus on the case $1<a_t<b_t$, which leads to $\mu_t<1/2$ and $\sigma_t<\mu_t$, and in particular the regime where $\mu_t$ is small, as in the main text.
When $\mu_t\ll1$, Eq.~(\ref{eq:muFromAB}) indicates that $a_t\ll b_t$, and then $\mu_t\simeq a_t/b_t$, $\sigma_t\simeq \sqrt{a_t}/b_t$, and $\sigma_t/\mu_t\simeq 1/\sqrt{a_t}$.

The behaviour of the SCE model defined in Eqs.~(\ref{eq:CreatorModel}) and (\ref{eq:OneDimLGS}) for the case $\beta_T>0$ and $\beta_N=0$ can be understood from the following analysis.
Eq.~(\ref{eq:OneDimLGS}) yields the following equations.
\begin{equation}
a_{t+1}-1=(1+\beta_T)(a_t-1)
\label{eq:atTr}
\end{equation}
\begin{equation}
b_{t+1}-1=(1+\beta_T)(b_t-1)
\label{eq:btTr}
\end{equation}
The fact that $\sigma_{t+1}<\sigma_t$, which is intuitively trivial, can be formally checked by differentiating the following quantity with respect to $\beta_T$:
\begin{equation}
h(\beta_T)=\sigma^2_{t+1}=
\frac{[a_t+\beta_T(a_t-1)][b_t+\beta_T(b_t-1)]}{[a_t+b_t+\beta_T(a_t+b_t-2)]^2[a_t+b_t+\beta_T(a_t+b_t-2)+1]},
\label{eq:sigmaUpdateNoveltyOnly}
\end{equation}
where we have used Eq.~(\ref{eq:sigmaFromAB}).
We can then show $\partial h/\partial\beta_T<0$ for $\beta_T>0$.
By noting that the transformation in Eqs.~(\ref{eq:atTr}) and (\ref{eq:btTr}) for any finite $\beta_T$ can be realized by iterating infinitesimal transformations, it has been shown that $\sigma_{t+1}<\sigma_t$.
By recursively applying Eqs.~(\ref{eq:atTr}) and (\ref{eq:btTr}) and substituting the result into Eq.~(\ref{eq:sigmaFromAB}), we can also see that
\begin{equation}
\sigma_t\sim\frac{1}{(1+\beta_T)^{t/2}}\frac{(a_0-1)(b_0-1)}{(a_0+b_0-2)^3}\to0\quad(t\to\infty).
\end{equation}

We can show $\mu_{t+1}<\mu_t$ directly in a similar manner.
Alternatively, one can understand this by looking at the mode (peak position)
\begin{equation}
k_t=\frac{a_t-1}{a_t+b_t-2}.
\end{equation}
We can easily show that $k_t<\mu_t$ and $k_{t+1}=k_t$, which means that the peak position is invariant under the dynamics.
Since the difference $\mu_t-k_t$ decreases as the standard deviation $\sigma_t$ decreases, the result $\sigma_{t+1}<\sigma_t$ indicates $\mu_{t+1}<\mu_t$.
The mean $\mu_t$ will converge to the mode $k_t$ since $\sigma_t\to0$ as $t\to\infty$.

For the case of $\beta_N>0$ and $\beta_T=0$, we gave an intuitive argument in the main text that a slow manifold is formed where $\sigma/\mu$ is almost constant and slightly less than unity.
As discussed there, this slow manifold is formed because $\sigma/\mu$ is decreased by an update when it is close to unity, which is in turn due to the boundary at zero and the resulting asymmetric shape of the beta distribution.
Here we provide some mathematical analyses to support this intuitive argument.
First, we show that the heights of the two peaks of $\tilde{\phi}_t=\phi_te^{-\beta_N\phi_t}$ are equal.
The position of the right and left peaks (denoted by $\theta_+$ and $\theta_-$) are obtained by solving the following equation:
\begin{equation}
0=\frac{\partial\tilde{\phi}_t(\theta)}{\partial\theta}=\frac{\partial\phi_t(\theta)}{\partial\theta}e^{-\beta_N\phi_t(\theta)}(1-\beta_N\phi_t(\theta)).
\end{equation}
This yields $\phi_t(\theta_\pm)=1/\beta_N$.
Substituting this back into $\tilde{\phi}_t$, we have $\tilde{\phi}_t(\theta_\pm)=1/(e\beta_N)$, which is the height of the peaks.
Thus the contributions of the two peaks for the determination of $\phi_{t+1}$ are characterized by their position (mean) and width.
As can be seen in Fig.~\ref{fig:2}(e), when $\sigma/\mu$ is close to unity the width of the left peak is so much less than that of the right peak because of the boundary at zero that the $\phi_{t+1}$ is determined dominantly by the right peak.

Next, for the regime of parameters of our interest ($\mu_t\ll1$), $\sigma_t/\mu_t\simeq1/\sqrt{a_t}$ holds.
This means that the ratio $\sigma_t/\mu_t$ is smaller if $a_t$ is larger.
That is, the gradient of the beta distribution near zero is smaller.
As we see in Fig.~\ref{fig:2}(e), when the next-generation creator's model $\phi_{t+1}$ is dominantly determined by the right peak of $\tilde{\phi}_t$, $a_{t+1}>a_t$ generally holds.
This shows that $\sigma_{t+1}/\mu_{t+1}<\sigma_t/\mu_t$ when $\sigma_t/\mu_t$ is close to unity.
Moreover, one finds from the relation $\tilde{\phi}_t(\theta_\pm)=1/(e\beta_N)$ that $\theta_+$ increase as $\beta_N$ becomes larger, which in turn indicates that $a_t$ becomes larger.
Thus, for larger $\beta_N$, the asymptotic value of $\sigma_t/\mu_t$ tends to be smaller.

For the case both $\beta_N,\beta_T>0$, a slow manifold is formed in the regime where $\mu_t$ and $\sigma_t$ are small.
To understand this, first note that $a_t\ll b_t$, $\mu_t\simeq a_t/b_t$, and $\sigma_t\simeq \sqrt{a_t}/b_t$ hold for $\mu_t\ll1$.
The effect of the typicality term can be written as $a_{t+1}=a_t+\beta_T(a_t-1)$ and $b_{t+1}\simeq (1+\beta_T)b_t$ from Eqs.~(\ref{eq:atTr}) and (\ref{eq:btTr}), which indicates the following.
\begin{equation}
\frac{\sigma_t}{\mu_t}\ll1~\Rightarrow~\mu_{t+1}\simeq\mu_t,\quad\sigma_{t+1}\simeq\frac{1}{\sqrt{1+\beta_T}}\sigma_t.
\label{eq:TypEffectCase1}
\end{equation}
\begin{equation}
\frac{\sigma_t}{\mu_t}\simeq1~\Rightarrow~\mu_{t+1}\simeq\frac{1}{1+\beta_T}\mu_t,\quad\sigma_{t+1}\simeq\frac{1}{1+\beta_T}\sigma_t.
\label{eq:TypEffectCase2}
\end{equation}
This means that if $\sigma_t/\mu_t\ll1$ the typicality term decreases this ratio by a constant factor of $1/\sqrt{1+\beta_T}$.
On the other hand, the novelty term increases it by a factor that becomes larger for smaller $\sigma_t$ (one can confirm this for example in Fig.~\ref{fig:2}(b)).
Thus, the effect of the novelty term dominates over that of the typicality term for sufficiently small $\sigma_t$, in which case $\sigma_t/\mu_t$ increases.
When $\sigma_t/\mu_t\simeq1$, the effect of the typicality term is negligible, so the slow manifold is formed due to the effect of the novelty term.
On the slow manifold, the typicality term acts on the mean $\mu_t$ as in Eq.~(\ref{eq:TypEffectCase2}), which has the effect of a constant reduction factor, and the novelty term acts as illustrated in Fig.~\ref{fig:2}(g), which has a larger effect for smaller $\mu_t$.
Thus, for sufficiently small $\mu_t$, the dynamics is dominated by the novelty term, leading to a nearly exponential growth on the slow manifold.

The dynamics of the log-potential model can be solved as follows.
Substituting Eq.~(\ref{eq:LogNormal}) and $R_t={\rm ln}\,\theta$ into Eq.~(\ref{eq:Update}), we have
\begin{equation}
\tilde{\phi}_t(\theta)=\frac{1}{\sqrt{2\pi}\tilde{\sigma}_t\theta}{\rm exp}\bigg[-\frac{1}{2\tilde{\sigma}^2_t}\big\{{\rm ln}\,\theta-(\tilde{\mu}_t+\beta\tilde{\sigma}^2_t)\big\}^2+{\rm const}\bigg].
\end{equation}
By solving for the parameters of $\phi_{t+1}$ with the same mean and standard deviation, we find that
\begin{equation}
\tilde{\sigma}_{t+1}=\tilde{\sigma}_t=:\tilde{\sigma},
\end{equation}
\begin{equation}
\tilde{\mu}_{t+1}=\tilde{\mu}_t+\beta\tilde{\sigma}^2.
\end{equation}
Using the relations between $(\tilde{\mu}_t,\tilde{\sigma}_t)$ and $(\mu_t,\sigma_t)$ given in the main text, we obtain
\begin{equation}
\sigma_{t+1}/\mu_{t+1}=\sigma_t/\mu_t,
\end{equation}
\begin{equation}
\mu_{t+1}=\mu_te^{\beta\tilde{\sigma}^2},
\end{equation}
as claimed in the main text.

\section*{Additional Information}

The authors declare no competing interests.

\section*{Acknowledgment}

The authors would like to thank Masahiko Ueda and Nobuto Takeuchi for useful discussions and Kazuyoshi Yoshii for sharing the enka music data. This work was in part supported by Grant-in-Aid for Scientific Research on Innovative Areas No.~17H06386 from the Ministry of Education, Culture, Sports, Science and Technology (MEXT) of Japan and Grants-in-Aid for Scientific Research Nos.~16J05486, 16H02917, 16K00501, and 19K20340 from Japan Society for the Promotion of Science (JSPS). The work of E.N.\ was supported by the JSPS Postdoctoral Research Fellowship. 

\section*{Data Availability Statement}

The datasets generated and analyzed during the current study are available in the GitHub repository, \url{https://evomusstyle.github.io/}.
See also Data Analysis section above.

\section*{Author Contribution}

Both authors designed the work and wrote the main manuscript text.
EN conducted numerial experiments and prepared all the figures.
Both authors reviewed the manuscript.

\newpage

\renewcommand{\thefootnote}{\fnsymbol{footnote}}

\begin{center}
\begin{spacing}{1.5}
{\Large\bf\color{darkred}
Supplemental Material for\\
``Statistical Evolutionary Laws in Music Styles''
}
\end{spacing}

\vskip 1.2cm

Eita Nakamura$^1$\footnote[1]{Electronic address: \tt{eita.nakamura@i.kyoto-u.ac.jp}}
and Kunihiko Kaneko$^2$\footnote[2]{Electronic address: \tt{kaneko@complex.c.u-tokyo.ac.jp}}

\vskip 0.4cm

{\it
$^1$ The Hakubi Center for Advanced Research and Graduate School of Informatics,\\ Kyoto University, Sakyo, Kyoto 606-8501, Japan\\
$^2$ Department of Basic Science, University of Tokyo, Meguro, Tokyo 153-8902, Japan
}

\end{center}
\vskip 1.2cm

\renewcommand{\thefootnote}{\arabic{footnote}}

\tableofcontents


\section{Analysis on the Classical Music Data}
\label{sec:DataAnalysis}

\begin{figure}[t]
\centering
{\includegraphics[clip,width=0.65\columnwidth]{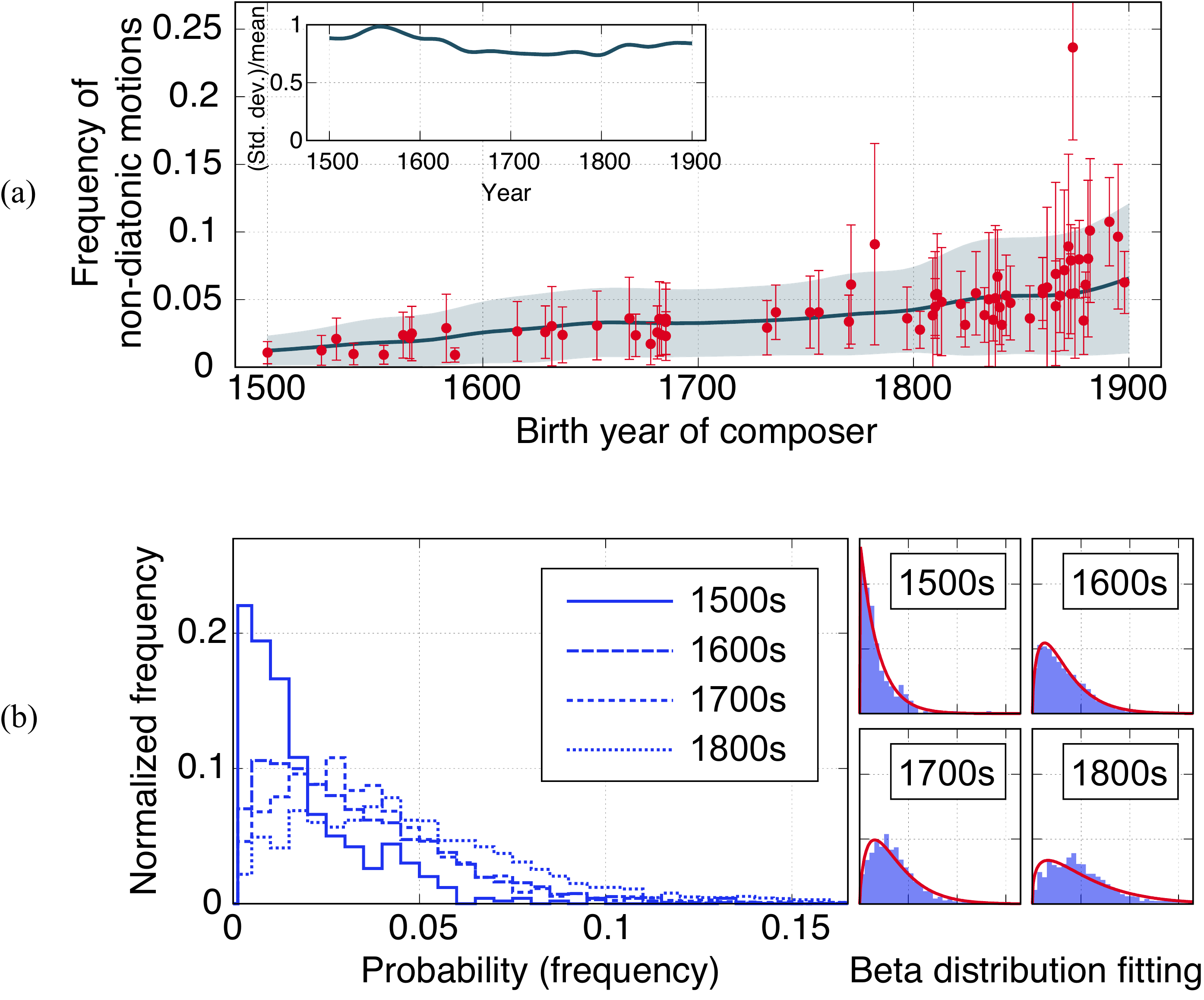}}
\vspace{-2mm}
\caption{Evolution of the distribution of frequencies of non-diatonic motions observed in Western classical music data. In (a), points and bars indicate the mean and standard deviation for each composer, and the step line and shade indicate those for time windows of 100-year width shifted in units of 25 years (spline interpolation applied). In (b), distributions obtained for each century.}
\label{fig:NondiatonicMotions}
\end{figure}

We can analyze frequencies of non-diatonic motions in the same way as for frequencies of tritones (Fig.~1 in the main text).
The result is shown in Fig.~\ref{fig:NondiatonicMotions}.
We can find the same statistical tendencies that are found for the frequencies of tritones, even though they are less clear:
\begin{itemize}\setlength\itemsep{-0.2em}
\item Beta-like distribution of frequency features
\item Steady increase of the mean and standard deviation
\item Nearly constant ratio of the mean and standard deviation
\end{itemize}

We used the birth year of the composer as the reference time of each musical piece in Fig.~1 in the main text and Fig.~\ref{fig:NondiatonicMotions} in this Supplemental Material.
This is because the composition year for each individual piece is not given in the dataset used.
Alternatively, if we use as the reference time the death year, the middle year (defined as the average of the birth and death years), and the active year (defined as the birth year plus 35 years to represent the active time of the composer's career) of the corresponding composer, we obtain similar results apart from shifts in time.

\begin{figure}[t]
\centering
{\includegraphics[clip,width=0.65\columnwidth]{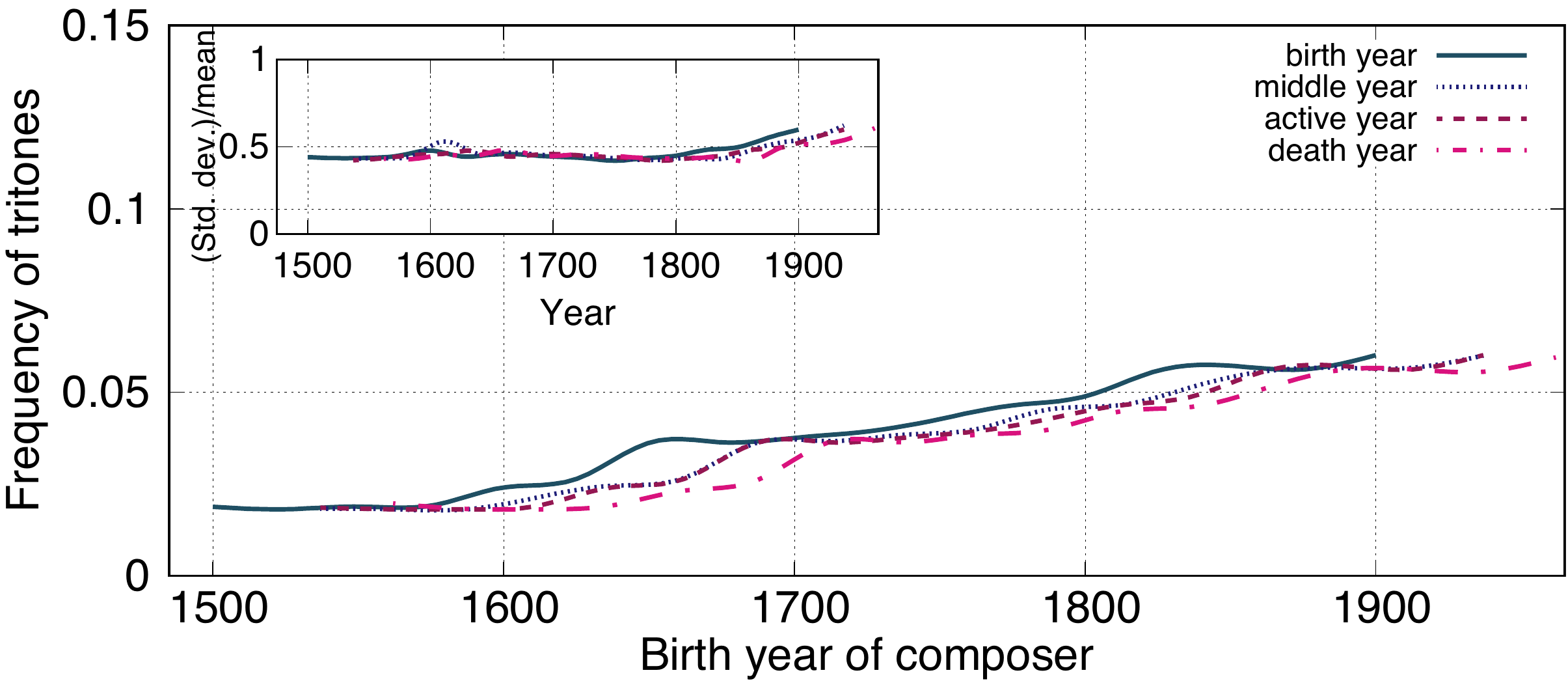}}
\vspace{-2mm}
\caption{Effect of using different reference times for the Western classical music data. Evolutions of the mean and standard deviation of frequencies of tritones, corresponding to Fig.~1(a) in the main text, are illustrated.}
\label{fig:GrowthTritoneFineDifferentYears}
\end{figure}

\newpage
\section{Analysis of the SCE models for other distributions}
\label{sec:ModelAnalysisOther}

%
\begin{figure}[t]
\centering
{\includegraphics[clip,width=0.45\columnwidth]{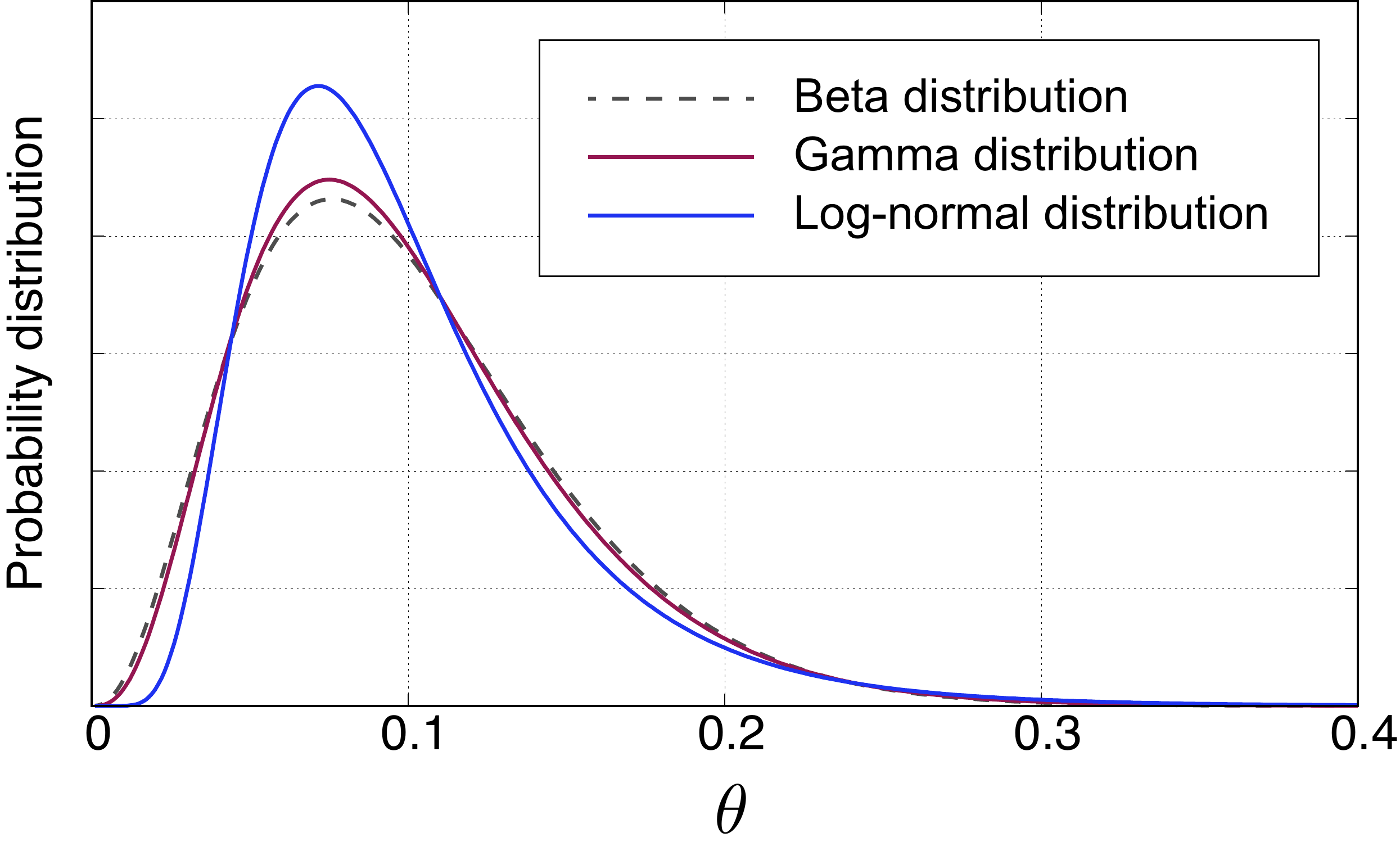}}
\vspace{-2mm}
\caption{Beta, gamma, and log-normal distribution for the same mean and standard deviation ($\mu=0.1$ and $\sigma=0.05$).}
\label{fig:DistributionComparison}
\end{figure}
\begin{figure}[t]
\centering
{\includegraphics[clip,width=0.7\columnwidth]{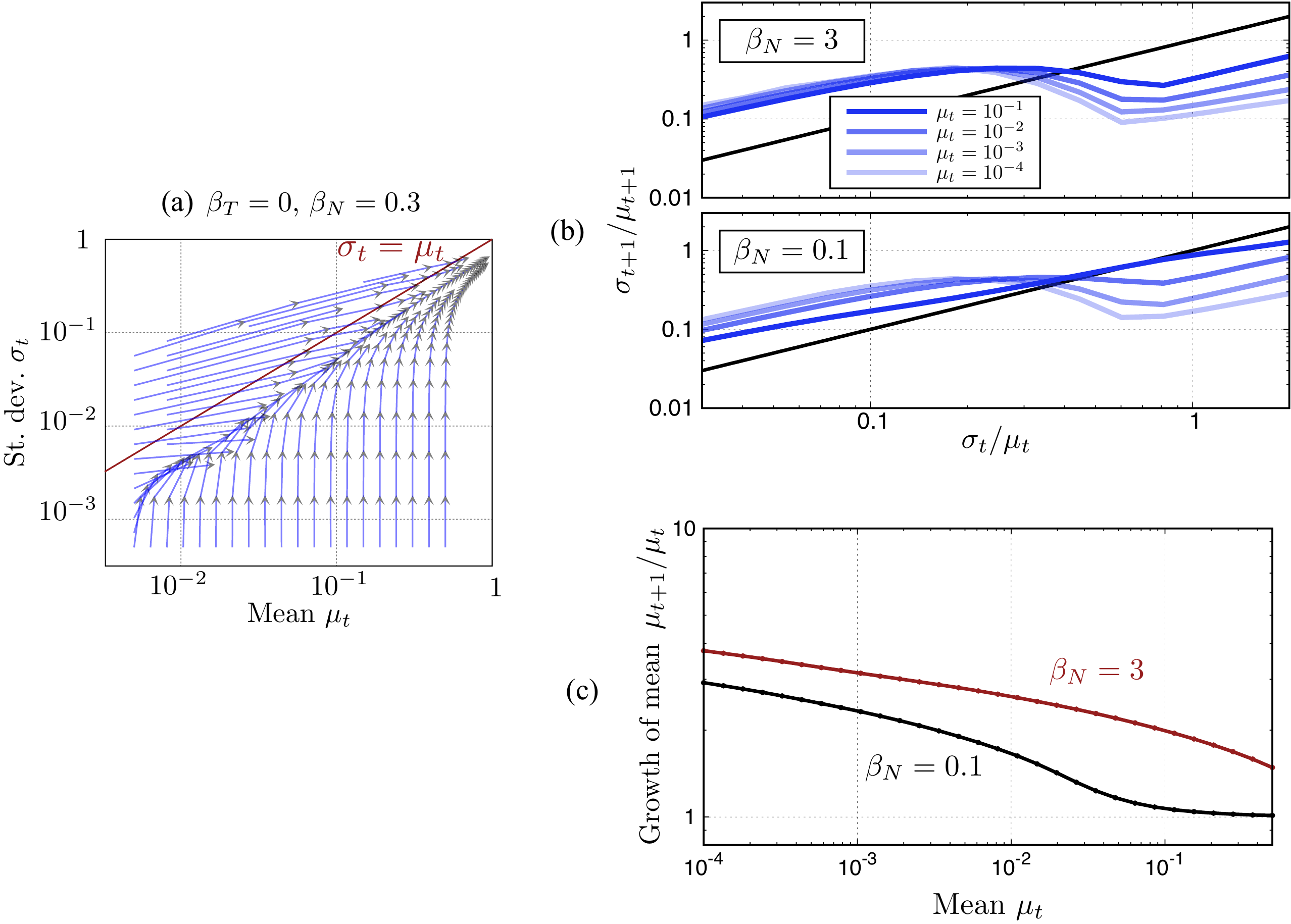}}
\vspace{-2mm}
\caption{Numerical analysis of the SCE model defined with the gamma distribution for the case $\beta_N>0$ and $\beta_T=0$. (a) Orbits of the SCE model  (b) Dynamics of the ratio $\sigma_t/\mu_t$. (c) Growth of the mean around the slow manifold ($\sigma_t/\mu_t=0.4$).}
\label{fig:GammaSCEMAnalysis}
\end{figure}
In the main text, we analyze the SCE model for the beta distribution.
Here we analyze the SCE models defined with the gamma and log-normal distributions to show the generality of the model analysis result, especially the existence of a slow manifold when the novelty term is active.
A gamma distribution is defined as
\begin{equation}
\phi_t(\theta)={\rm Gamma}(\theta;a_t,b_t)=\frac{b_t^{-a_t}}{\Gamma(a_t)}x^{a_t-1}e^{-x/b_t},
\label{eq:GammaDistribution}
\end{equation}
and the parameters $a_t,b_t>0$ are related to the mean and standard deviation as
\begin{equation}
\mu_t=a_tb_t,\quad \sigma_t=b_t\sqrt{a_t}.
\end{equation}
A log-normal distribution is defined as
\begin{equation}
\phi_t(\theta)={\rm LN}(\theta;\tilde{\mu}_t,\tilde{\sigma}_t)=\frac{1}{\sqrt{2\pi}\tilde{\sigma}_t\theta}{\rm exp}\bigg[-\frac{({\rm ln}\,\theta-\tilde{\mu}_t)^2}{2\tilde{\sigma}^2_t}\bigg],
\label{eq:LogNormalDistribution}
\end{equation}
and the parameters $\tilde{\mu}_t\in(0,\infty)$ and $\tilde{\sigma}_t>0$ are related to the mean and standard deviation as
\begin{equation}
\mu_t={\rm exp}(\tilde{\mu}_t+\tilde{\sigma}^2_t/2),\quad
\sigma_t/\mu_t=\sqrt{{\rm exp}(\tilde{\sigma}^2_t)-1}.
\end{equation}
Both of these probability distributions are defined in the range $\theta\in(0,\infty)$, so they are not strictly proper for the probability parameter $\theta$ restricted in the range $(0,1)$.
Nevertheless, when the mean $\mu_t$ is smaller than unity and the standard deviation is sufficiently small, the supports of these distributions are effectively bounded in the range $(0,1)$.
We study these distributions for the demonstration purpose.

\begin{figure}[tbh]
\centering
{\includegraphics[clip,width=0.7\columnwidth]{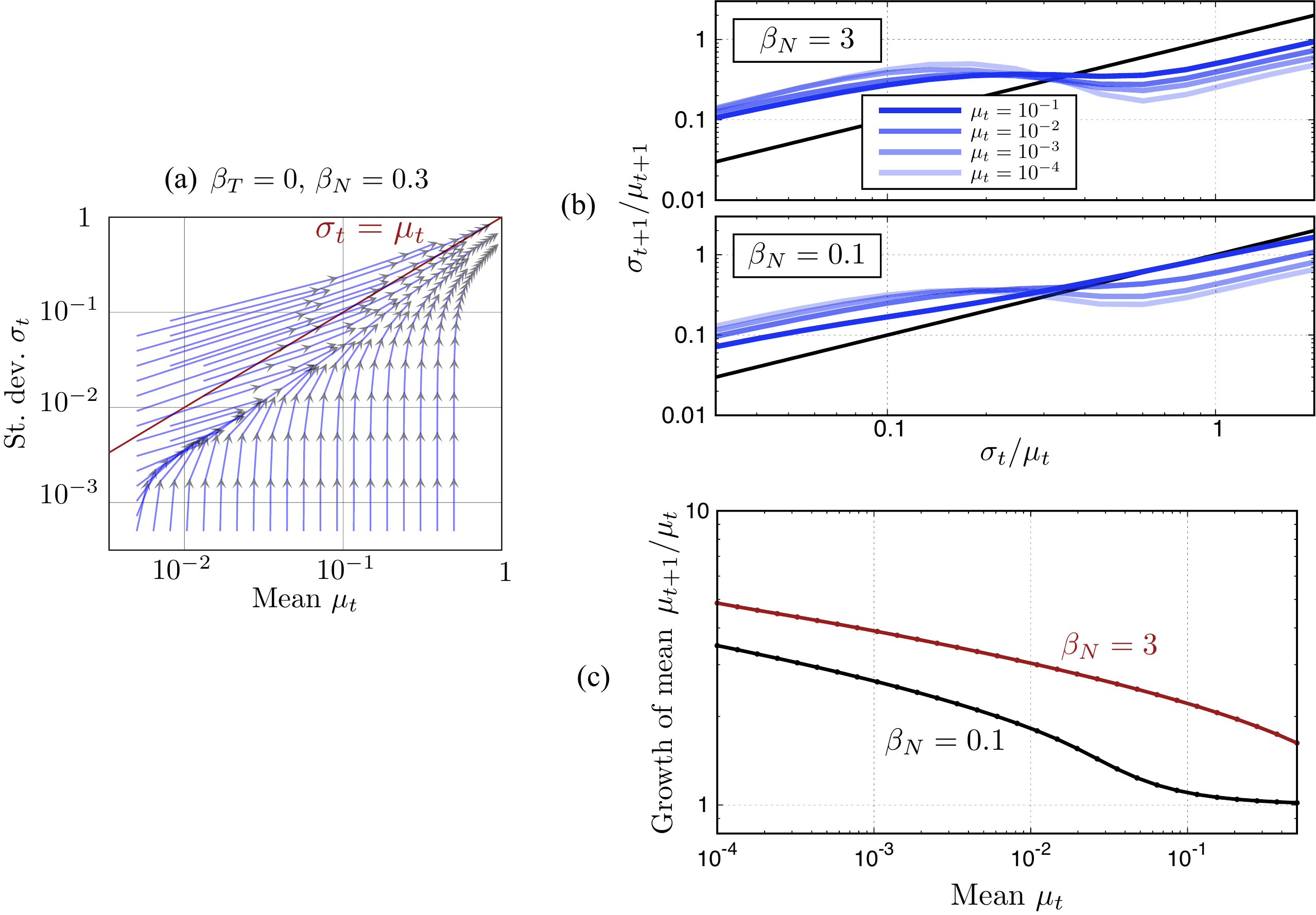}}
\vspace{-2mm}
\caption{Numerical analysis of the SCE model defined with the log-normal distribution for the case $\beta_N>0$ and $\beta_T=0$. (a) Orbits of the SCE model  (b) Dynamics of the ratio $\sigma_t/\mu_t$. (c) Growth of the mean around the slow manifold ($\sigma_t/\mu_t=0.4$).}
\label{fig:LogNormalSCEMAnalysis}
\end{figure}
The gamma, log-normal, and beta distributions are compared in Fig.~\ref{fig:DistributionComparison}, where distributions with the same mean and standard deviation ($\mu=0.1$ and $\sigma=0.05$) are shown.
As we see in the figure, the shapes of these three distributions are generally similar for a small mean and for a standard deviation smaller than the mean.

SCE models for the gamma and log-normal distributions are defined by substituting Eqs.~(\ref{eq:GammaDistribution}) and (\ref{eq:LogNormalDistribution}) into Eq.~(4) in the main text, respectively.
We can conduct numerical analyses similarly as in the main text.
Focusing on the case $\beta_N>0$ and $\beta_T=0$, results of numerical analyses are shown in Figs.~\ref{fig:GammaSCEMAnalysis} and \ref{fig:LogNormalSCEMAnalysis}.
From the results in the figures and with the same argument as in the main text, one can see a slow manifold in which $\sigma_t/\mu_t$ is kept almost constant and $\mu_t$ grows nearly exponentially, similarly as in the case of the beta distribution.

\newpage
\section{Additional Comparison between the SCE Model and the Log-Potential Model}
\label{sec:AdditionalComparison}

In Figs.~4(a) and 4(b) of the main text, we compared how the SCE model and the log-potential model can fit the real data of classical music.
There, the model parameters were optimized to best fit the two sets of data (frequencies of tritones and non-diatonic motions).
Here we report the results when the model parameters are fitted to the two sets of data individually.

\begin{figure}[t]
\centering
{\includegraphics[clip,width=0.9\columnwidth]{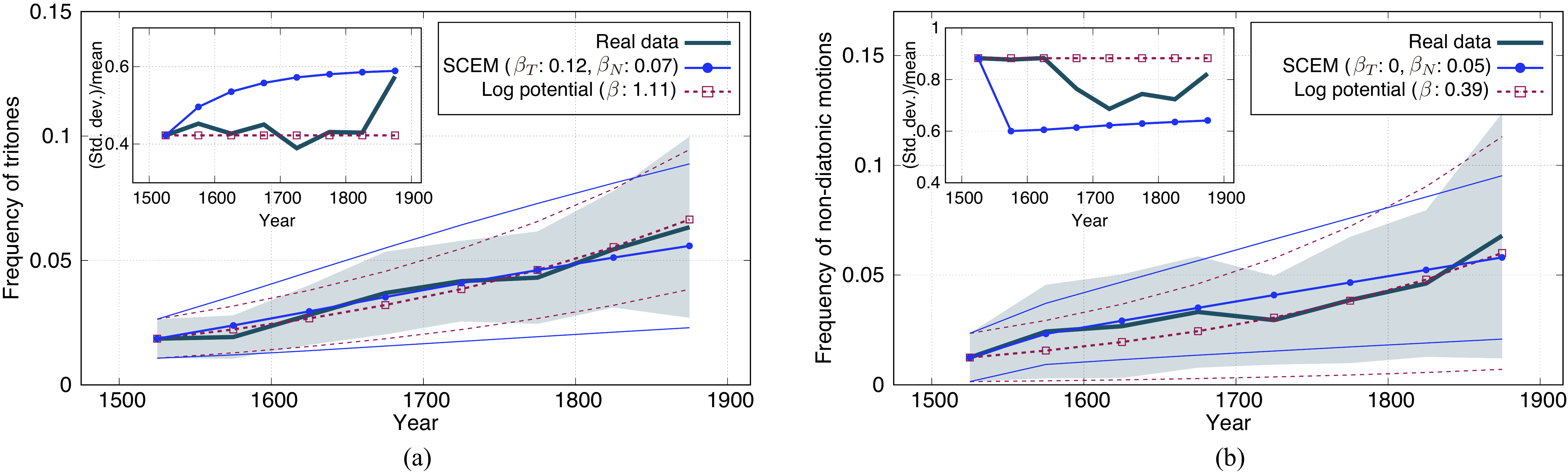}}
\caption{Comparisons between model predictions and real data. Bold lines indicate means, and thin lines and shadow indicate the ranges of $\pm1$ standard deviation. Model parameters are optimized individually to fit the two datasets to minimize the squared error of predicted means and standard deviations (optimal parameters are shown in the insets).
}
\label{fig:IndividualFit}
\end{figure}

The results are shown in Figs.~\ref{fig:IndividualFit}(a) and \ref{fig:IndividualFit}(b).
The root mean squared errors of the (tritone, non-diatonic motion) data are ($4.4\times10^{-3}$, $7.2\times10^{-3}$) for the SCE model and ($3.0\times10^{-3}$, $5.8\times10^{-3}$) for the log-potential model.
Compared to the results in the main text, these results show that for the SCE model the precision of the individual fit is similar to that of the simultaneous fit, and that the log-potential model can fit individual data slightly better than the SCE model.
This shows that although the log-potential model is flexible for fitting individual sets of data, it cannot fit both sets of data simultaneously, confirming that it is not trivial to fit both sets of data in a unified manner.

\end{document}